\documentclass[useAMS,usenatbib]{mn2e}
\usepackage{psfrag,graphicx}
\usepackage{color}
\title [Signatures of accretion and infall in the presence of line trapping]{Lyman alpha emission from the first galaxies: Signatures of accretion and infall in the presence of line trapping}
\author[M.A. Latif et al.]
{M.A.~Latif $^1$,
Dominik.R.G.Schleicher $^2$$^,$$^3$,
M.~Spaans$^1$,
S.~Zaroubi$^1$$^,$$^4$ \\
$^1$ Kapteyn Astronomical Institute, University of Groningen, P.O.~Box 800, 9700 AV Groningen, The Netherlands \\
$^2$ Leiden Observatory, Leiden University, P.O.~Box 9513, NL-2300 RA Leiden, The Netherlands \\
$^3$ ESO Garching, Karl-Schwarzchild-Str.2, 85748 Garching bei Munchen, Germany \\
$^4$ Physics Department, Technion, Haifa 32000, Israel}

\date{Accepted 2011 February 01. Received 2011 January 25; in original form 2010 December 09}

\pagerange{\pageref{firstpage}--\pageref{lastpage}} \pubyear{2009}

\def\LaTeX{L\kern-.36em\raise.3ex\hbox{a}\kern-.15em
    T\kern-.1667em\lower.7ex\hbox{E}\kern-.125emX}

\begin{document}

\bibliographystyle{mn2e}

\label{firstpage}

\maketitle

 \setlength{\bibhang}{2em}

\begin{abstract}
The formation of the first galaxies is accompanied by large accretion flows and virialization shocks, during which the gas is shock-heated to temperatures of $\sim10^4$~K, leading to potentially strong fluxes in the Lyman alpha line. Indeed, a number of Lyman alpha blobs has been detected at high redshift. In this letter, we explore the origin of such Lyman alpha emission using cosmological hydrodynamical simulations that include a detailed model of atomic hydrogen as a multi-level atom and the effects of line trapping with the adaptive mesh refinement code FLASH. We see that baryons fall into the center of a halo through cold streams of gas, giving rise to a Lyman alpha luminosity of at least $\rm 10^{44}~erg~s^{-1}$ at $\rm z=4.7$, similar to observed Lyman alpha blobs. We find that a Lyman alpha flux of $\rm 5.0\times 10^{-17} erg~cm^{-2}~s^{-1}$ emerges from the envelope of the halo rather than its center, where the photons are efficiently trapped. Such emission can be probed in detail with the upcoming James Webb Space Telescope (JWST) and will constitute an important probe of gas infall and accretion. 
\end{abstract}


\section{Introduction}
Early-type galaxies are known to produce copious Lyman alpha photons due to their spatially extended gas distribution. In protogalactic halos gas transfers its gravitational binding energy to the excitation of hydrogen atoms which results in Lyman alpha emission \citep{2000ApJ...537L...5H,2006ApJ...649...14D,2006ApJ...649...37D}. Therefore, the gas in dark matter halos exceeding a virial temperature of $\sim10^4$~K may be detected in the Lyman alpha line emission. The presence of ionizing radiation sources may enhance the Lyman alpha flux as gas is photo-ionized, and and supernova feedback may further increase the escape fraction by generating a more clumpy structure. These ionization sources could be the first stars or mini-quasars. A luminous quasar can boost the emission of Lyman alpha photons by several orders of magnitude \cite{2001ApJ...556...87H}.

Many Lyman alpha blobs (LABs) have been observed at high redshift \citep{2009ApJ...693.1579Y,2009ApJ...696.1164O,2004AJ....128..569M,2000ApJ...532..170S}. In the light of recent detections at redshift $\rm > 7$ \citep{2010Natur.467..940L,2010arXiv1011.5500V}, it is of high interest to understand what drives the emission and how it is spatially distributed. A number of LABs have been observed whose most probable origin is cold accretion of gas onto dark matter halos \citep{2006A&A...452L..23N,2007MNRAS.378L..49S}. 

Numerical simulations show that cooling by Lyman alpha radiation induces collapse in protogalactic halos. Moreover, baryons accumulate into the center of halos by penetration of cold streams of gas through the shock heated medium \citep{2001ApJ...562..605F,2008ApJ...682..745W,2009ApJ...696.1065J,2009MNRAS.396..343R,2009Natur.457..451D,2009MNRAS.395..160K,2010MNRAS.407..613G,2010MNRAS.tmp.1837L,2010MNRAS.402.1249S}. Cold streams with temperatures of $\sim10^4$~K could be potential sources of spatially extended Lyman alpha emission \citep{2009MNRAS.400.1109D}. \cite{2010MNRAS.tmp.1427J} found that Lyman alpha radiation can also be emitted during accretion of gas on black holes formed by direct collapse in the first galaxies. Such emission is potentially detectable with JWST\footnote{http://www.stsci.edu/jwst/instruments/nirspec/sensitivity}. Similarly, it allows to probe the starburst component through the enhanced emission in several recombination lines  \citep{2009MNRAS.399...37J}.

The presence of large columns of neutral hydrogen gas causes the trapping of Lyman alpha photons \citep{2006ApJ...652..902S,2010MNRAS.tmp.1837L,2010ApJ...712L..69S}, which was neglected in some of the previous studies. In this letter, we aim at a self-consistent modeling of Lyman alpha emission driven by accretion flows, including dynamics, non-equilibrium chemistry of H, $\rm H^{+}$, He, $\rm He^{+}$ and $\rm He^{++}$, detailed level populations of atomic hydrogen, as well as the trapping of hydrogen line photons due to large column densities. The prime objective of this work is to study the origin and spatial distribution of Lyman alpha emission, which is detectable with JWST.


\section {Modeling of the physics}
For our simulations, we employ an extended version of the adaptive mesh refinement (AMR) hydrodynamics code FLASH \citep{Dubey2009512}. FLASH is a module based Eulerian grid parallel simulations code. We use an AMR grid to achieve high dynamic resolution in the regions of interest. We employ the directionally split piece wise parabolic method (PPM) for hydrodynamic calculations, which is an improved form of the Godunov method \citep{1984JCoPh..54..174C}. This method is well suited for flows involving shocks and contact discontinuities. The dark matter is simulated based on the particle mesh (PM) method.
In order to create Gaussian random field initial conditions, we run the COSMICS package developed by \cite{1995astro.ph..6070B}. We start our simulation at redshift 100. Our computational periodic box has a comoving size of 10 Mpc. We perform our simulations in accordance to the $\rm \Lambda$CDM model with WMAP 5-years parameters. We select a $4\times 10^{9} M_{\odot}$ halo at redshift 7 and follow its collapse. We enforce 8 additional levels of refinement, which gives 15 levels of refinement in total. In this way, we obtain a dynamic resolution limit of 70pc (co-moving), and can still follow the further evolution until redshift $\rm z=4.7$. We impose the Truelove criterion by refining according to the Jeans length \citep{1997ApJ...489L.179T}. We resolve the Jeans length by at least 20 cells.Such resolution was shown to be sufficient for accurately resolving turbulent structures during
gravitational collapse (Federrath et al. 2010). When the highest refinement level is reached, we prevent artificial fragmentation via Jeans-heating. We assume a  primordial gas composition with $\rm 75 \%$ hydrogen and $\rm 25 \%$ helium by mass.

We have developed a chemical network for the non-equilibrium modeling of non-molecular species. For this purpose, we solve the rate equations of the following species $\rm H,H^{+},He,He^{+},He^{++}$ and $\rm e^{-}$ for non-equilibrium ionization. The rate equations for these species are solved using the backward difference formula (BDF) method of \cite{1997NewA....2..209A}. We adopt the chemical rates of \cite{1997NewA....2..181A} and \cite{2008A&A...490..521S}. We further solve for the level populations of atomic hydrogen up to the fifth electronically excited state, and model the non-equilibrium cooling including hydrogen line emission, collisional ionization cooling, recombination cooling, Bremsstrahlung cooling and Compton cooling/heating. The transition rates for the level populations are based on work of \citet{2001ApJ...546..635O}.
At an optical depth $\rm \tau_{0} > 10^{7}$, the photon escape time becomes longer than the gas free fall time. Because of the weak dependence of the photon escape time on the gas number density, $\rm t_{ph} \propto n^{-1/9}$, trapping becomes important important during the collapse since the latter scales as $\rm n^{-1/2}$. A detailed discussion of such line trapping effects is given by \citet{2001ApJ...546..635O, 2006ApJ...652..902S, 2010ApJ...712L..69S}. We have computed the Lyman alpha luminosity by determining the Lyman alpha emissivity and integrating it over the virial volume of a halo. Further details of the luminosity and flux calculation can be found in \cite{2006ApJ...649...14D}.

\begin{figure*}
\centering
\begin{tabular}{c c}
\begin{minipage}{8cm}
\includegraphics[scale=0.30]{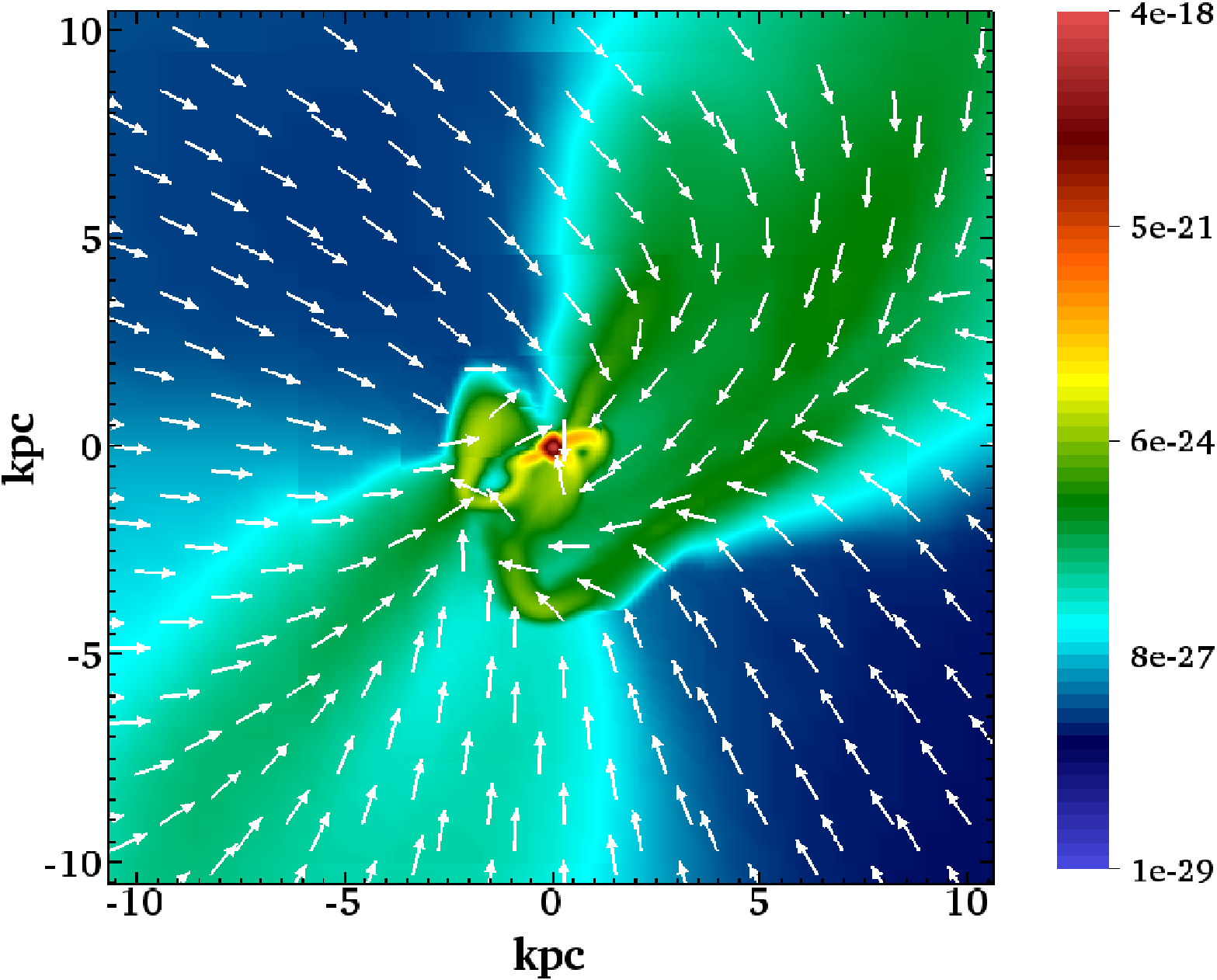}
\end{minipage} &
\begin{minipage}{8cm}
\includegraphics[scale=0.30]{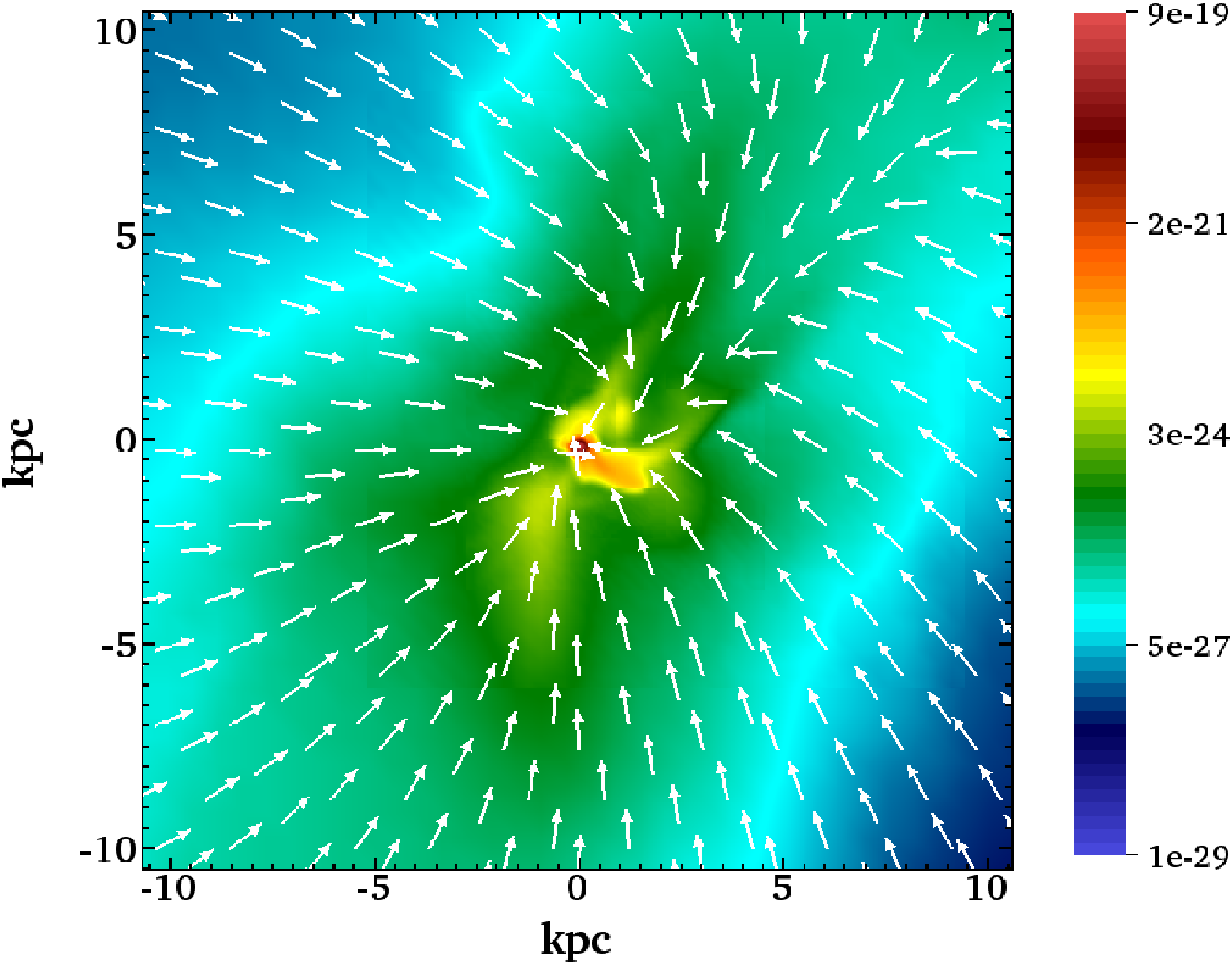}
\end{minipage} \\  \\

\begin{minipage}{8cm}
\includegraphics[scale=0.30]{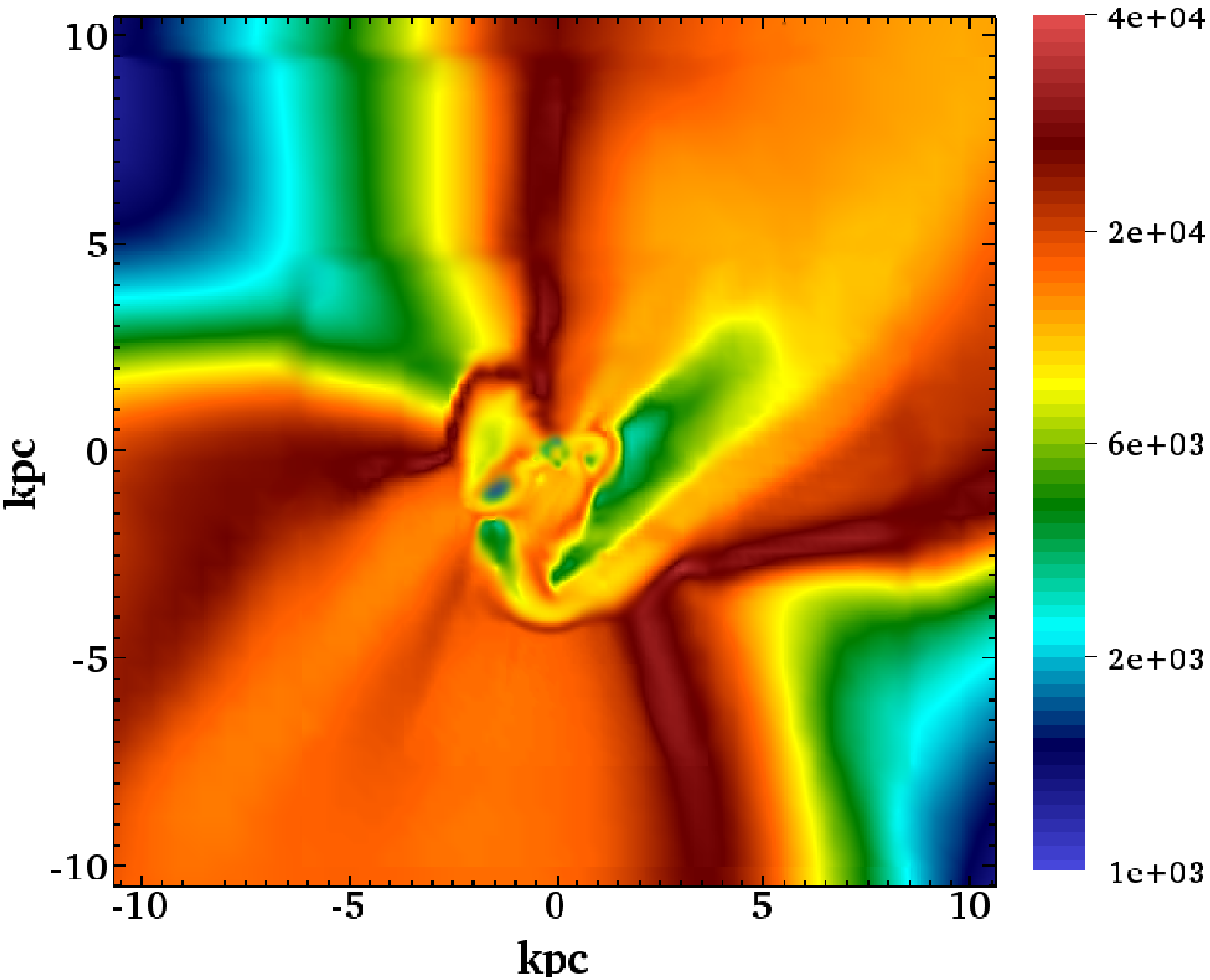}
\end{minipage} &

\begin{minipage}{8cm}
\includegraphics[scale=0.30]{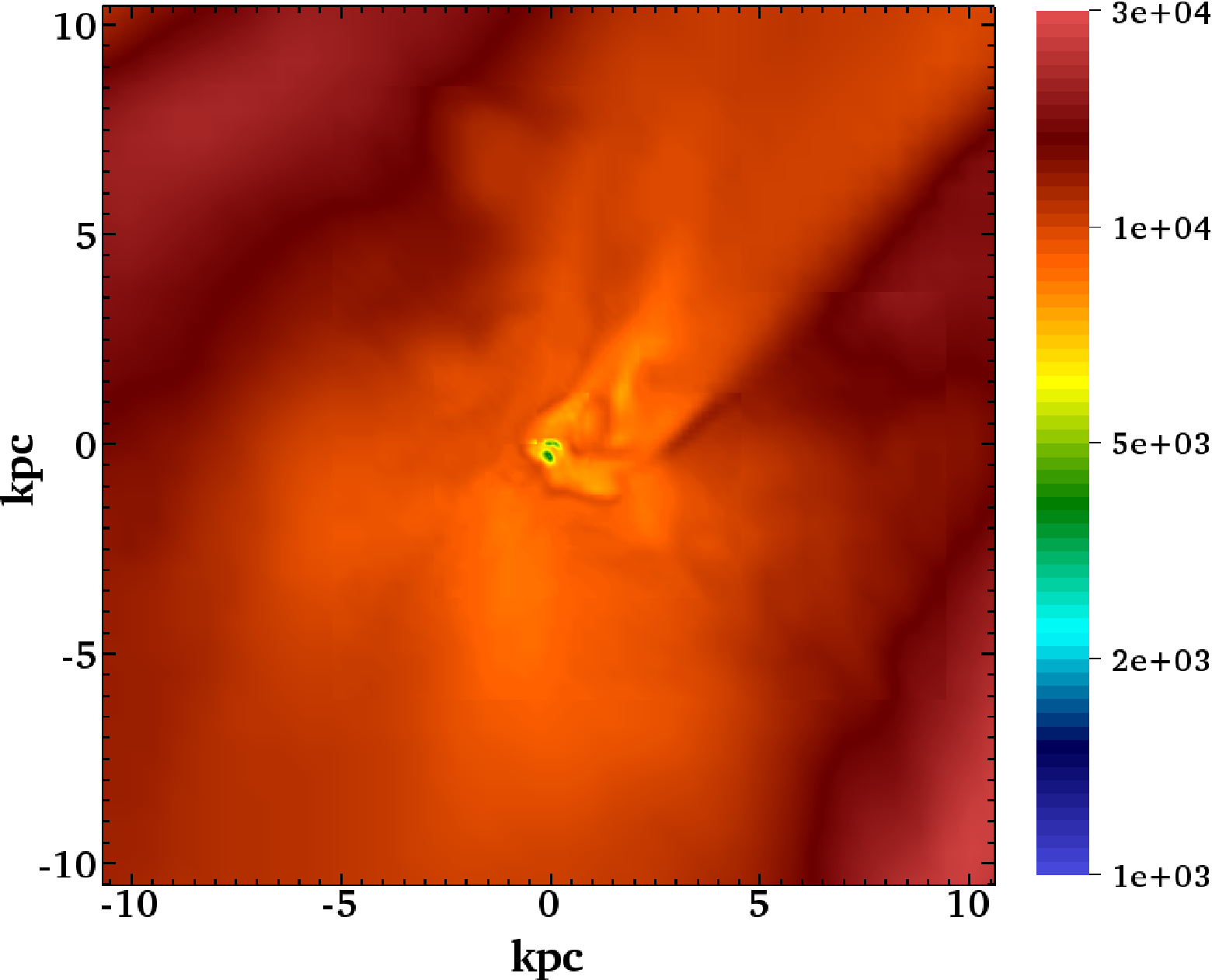}
\end{minipage}

\end{tabular}
\caption{The upper panels of this figure show the density slices through the center of the halo. The left panel is for redshift 4.7 and right panel for redshift 8. Velocity vectors are overplotted on the density slices. The bottom two panels show the corresponding temperatures. The figure shows the inner region of 20 kpc in comoving units.}
\label{figure1}
\end{figure*}

\section{Results and Conclusions}
\begin{figure*}
\centering
\begin{tabular}{c c}
 \begin{minipage}{8cm}
\includegraphics[scale=0.23]{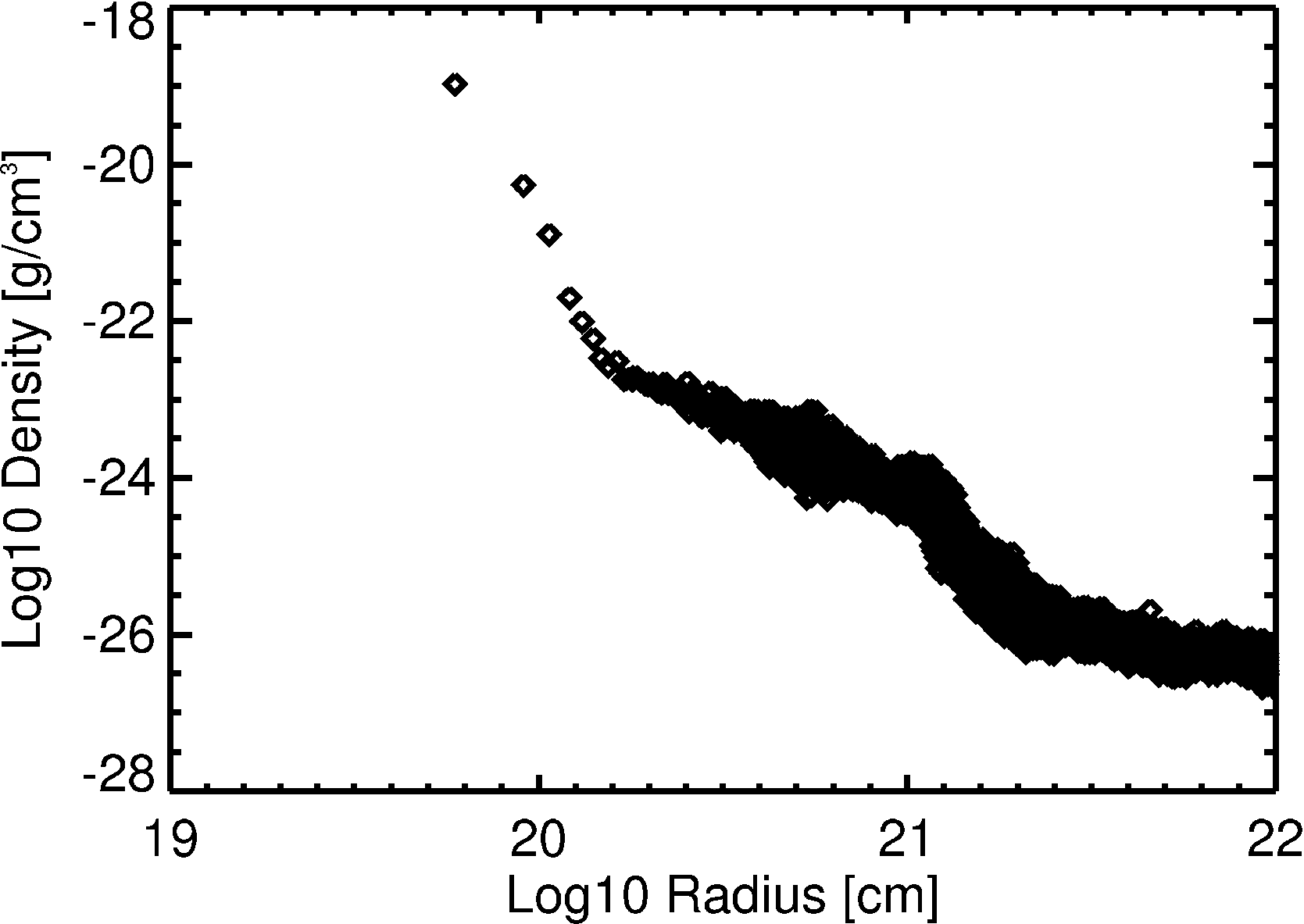}
\end{minipage} &

\begin{minipage}{8cm}
\includegraphics[scale=0.23]{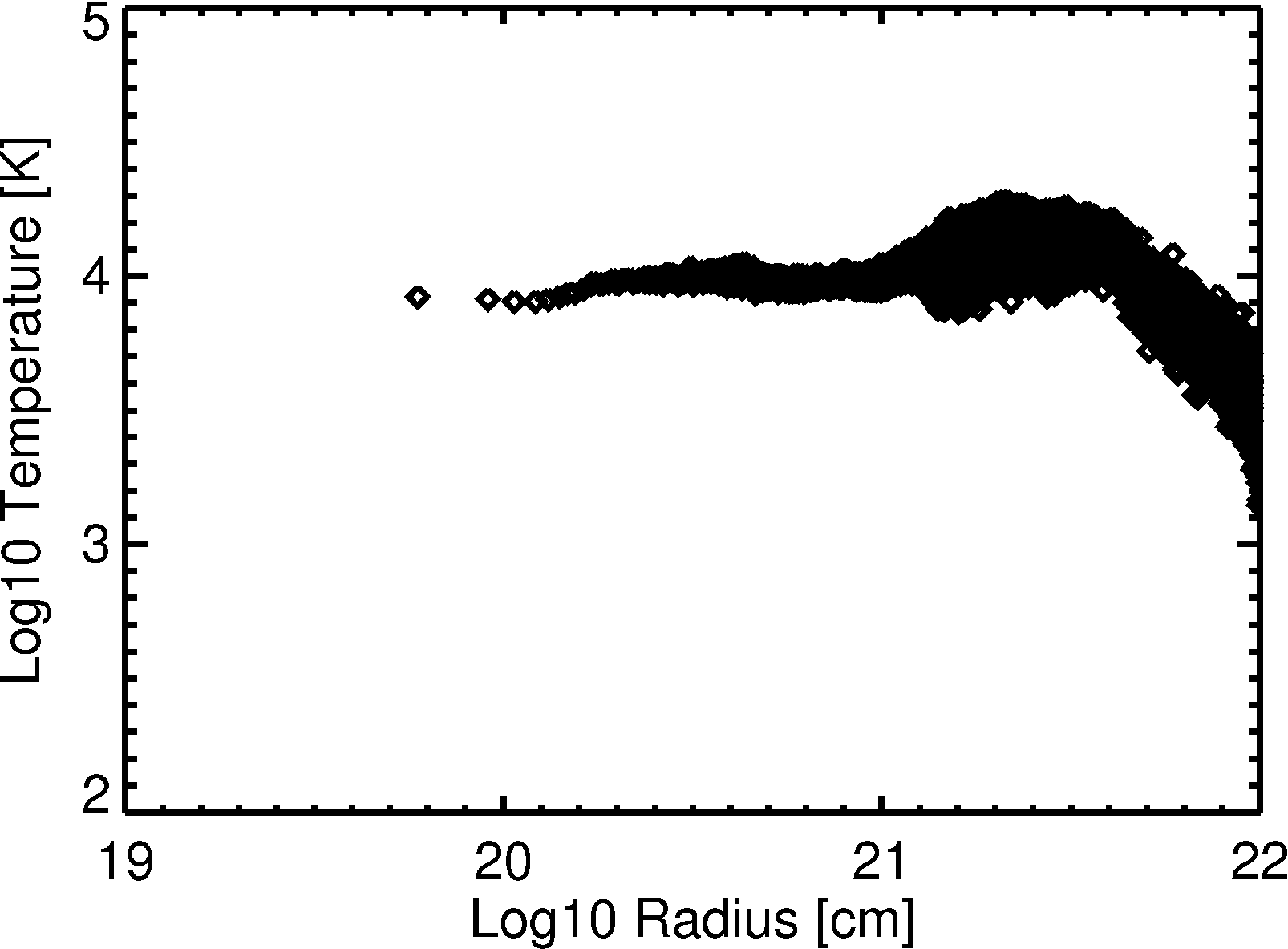}
\end{minipage} \\  \\

\begin{minipage}{8cm}
\includegraphics[scale=0.23]{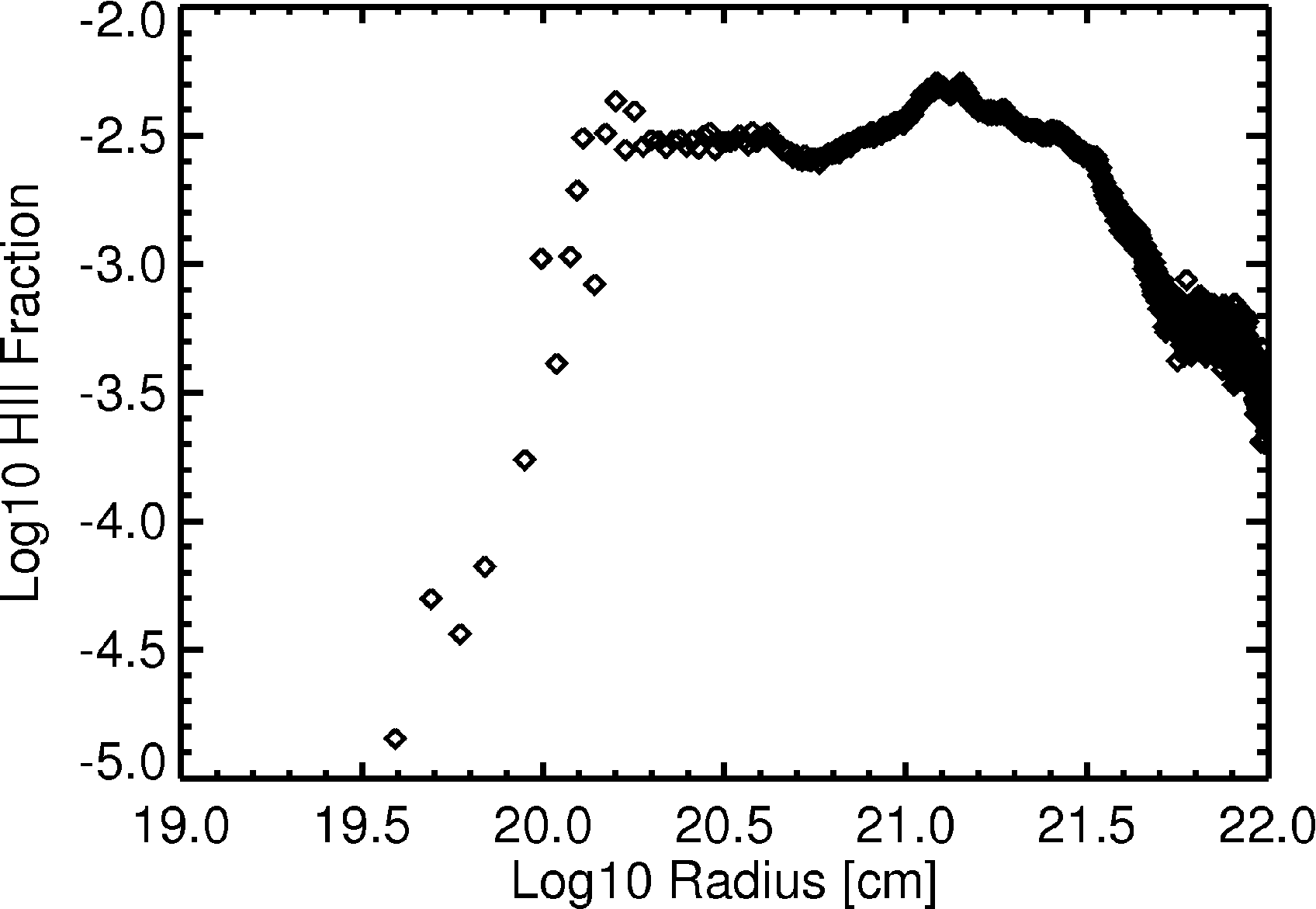}
\end{minipage} &

\begin{minipage}{8cm}
\includegraphics[scale=0.22]{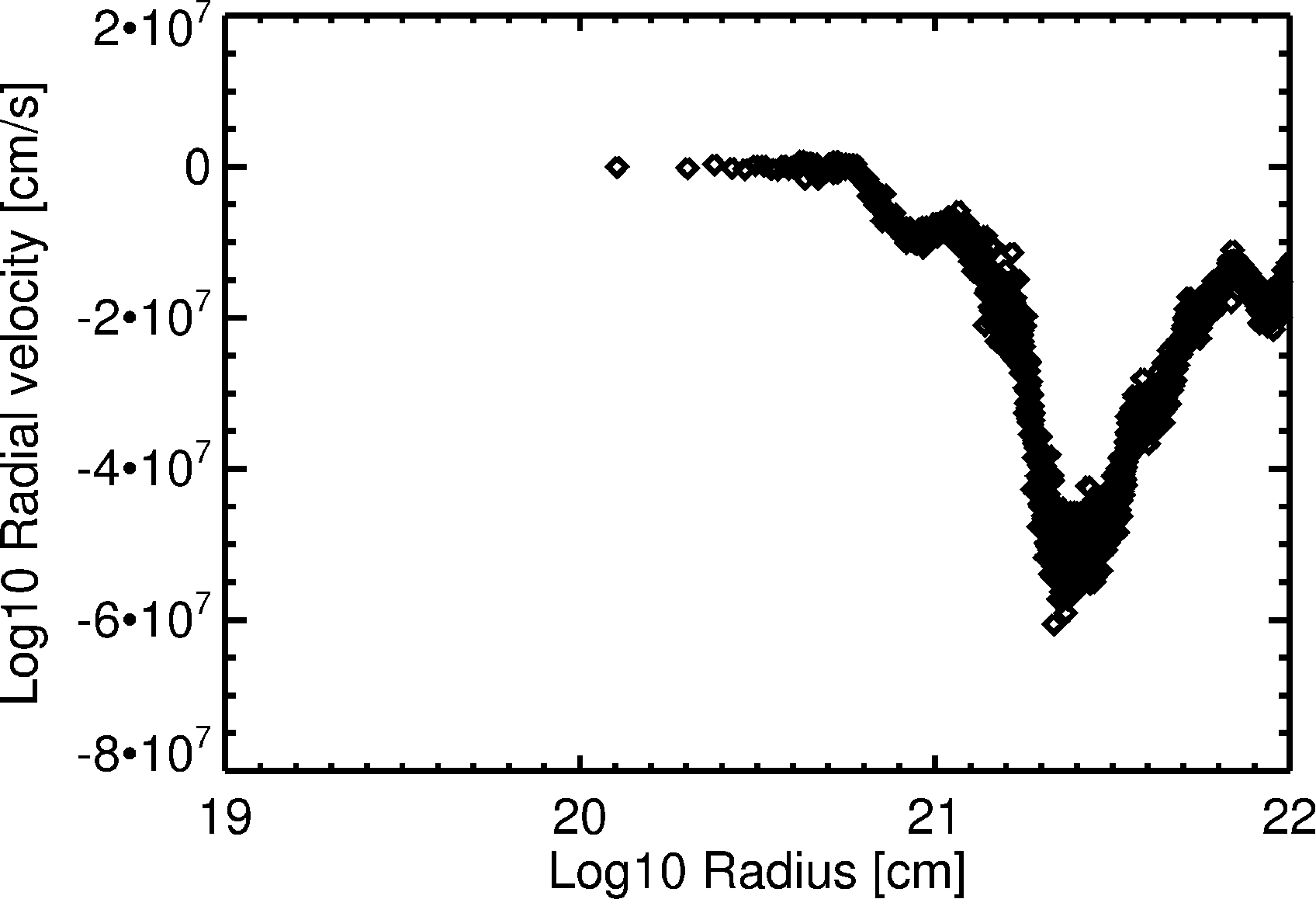}
\end{minipage}

\end{tabular}
\caption{All panels in this figure are log log plots. The upper left panel shows the radial density profile of the halo. The radial temperature profile for the halo is depicted in the upper right panel. HII abundance is shown in the lower left panel. The radial velocity of the halo is depicted in the lower right panel.}
\label{figure2}
\end{figure*}

\begin{figure*}
\centering
\begin{tabular}{c c}
 \begin{minipage}{8cm}
\includegraphics[scale=0.23]{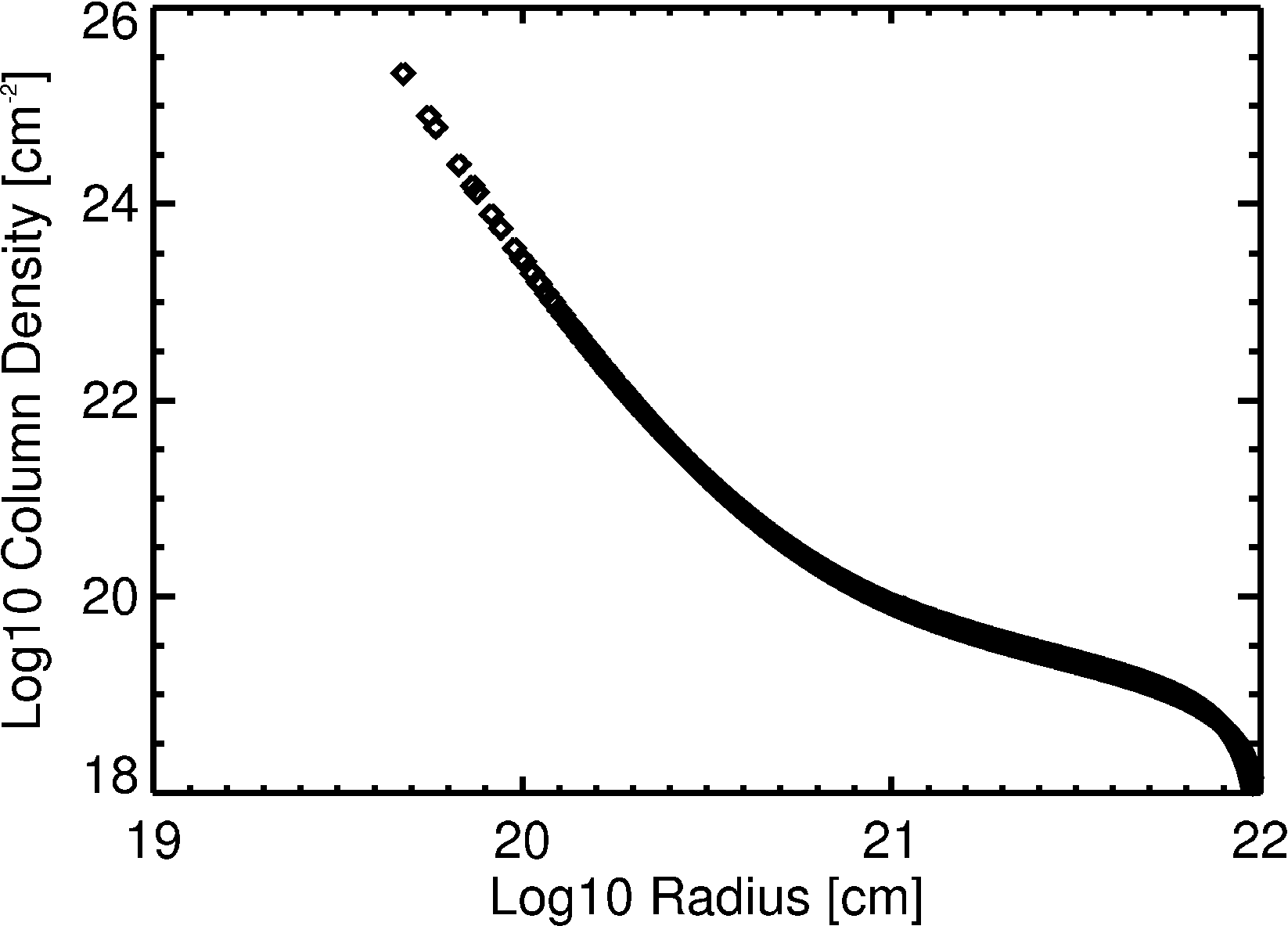}
\end{minipage} &

\begin{minipage}{8cm}
\includegraphics[scale=0.23]{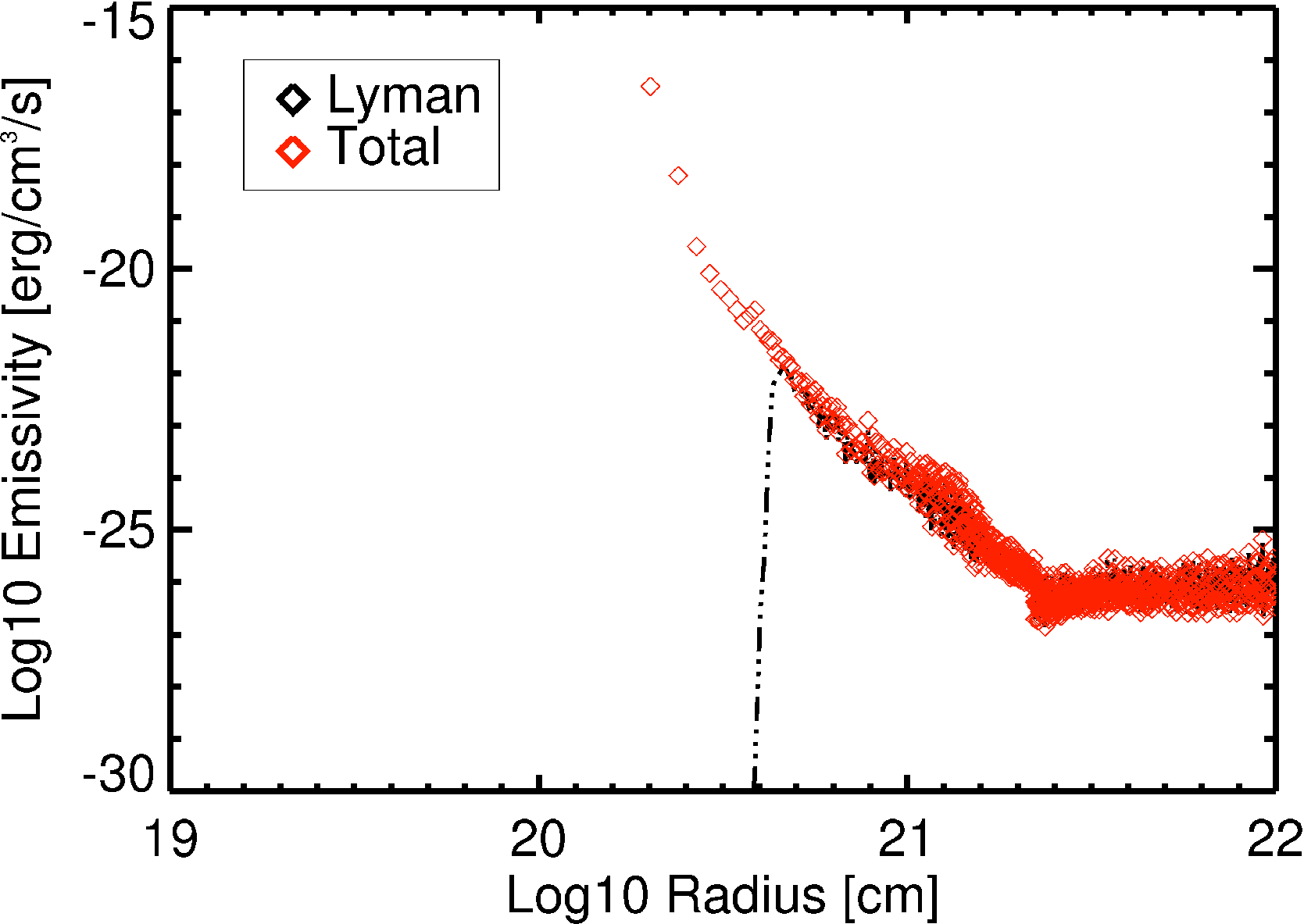}
\end{minipage} \\  \\

\begin{minipage}{8cm}
\includegraphics[scale=0.23]{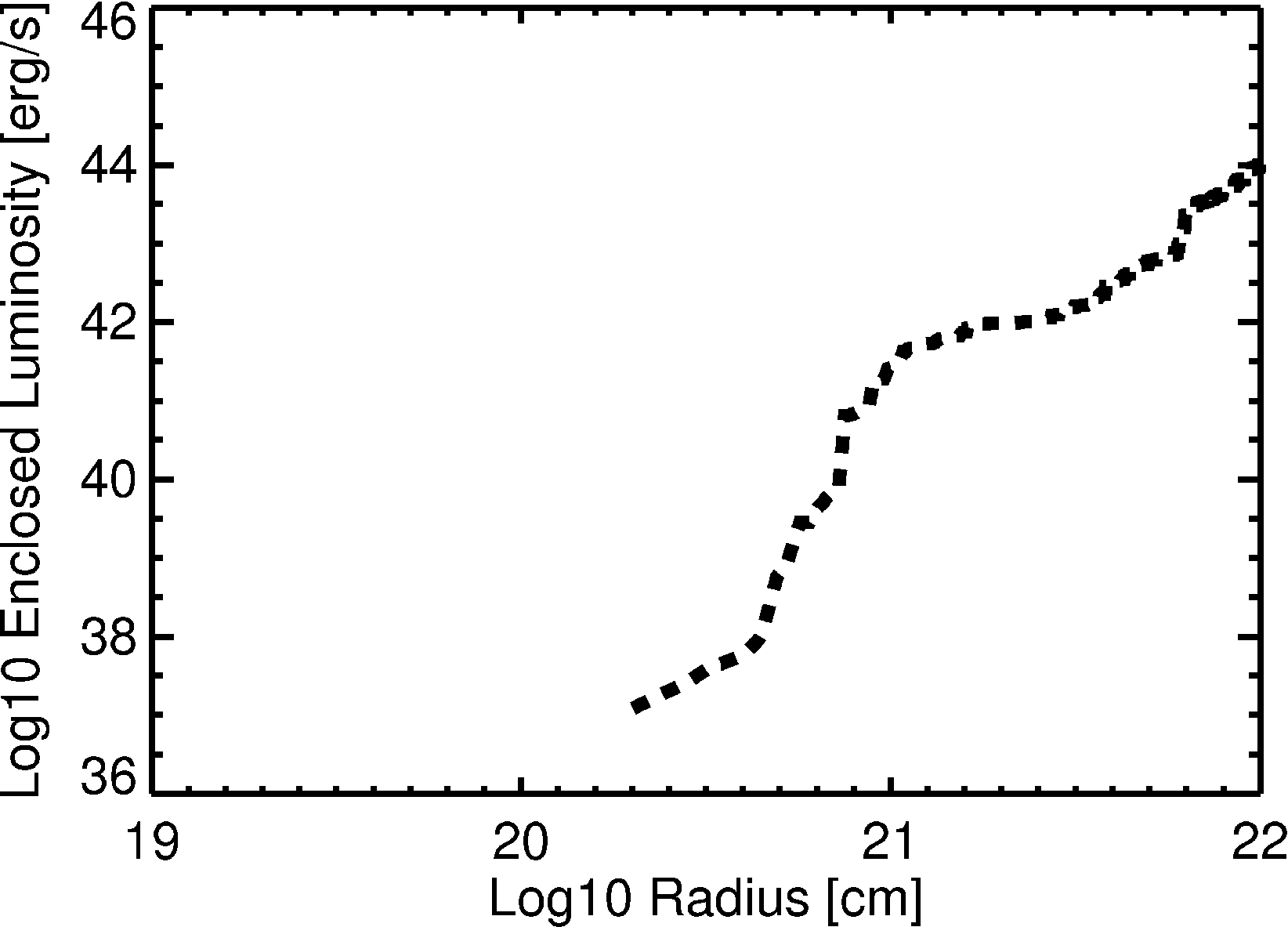}
\end{minipage} &

\begin{minipage}{8cm}
\includegraphics[scale=0.23]{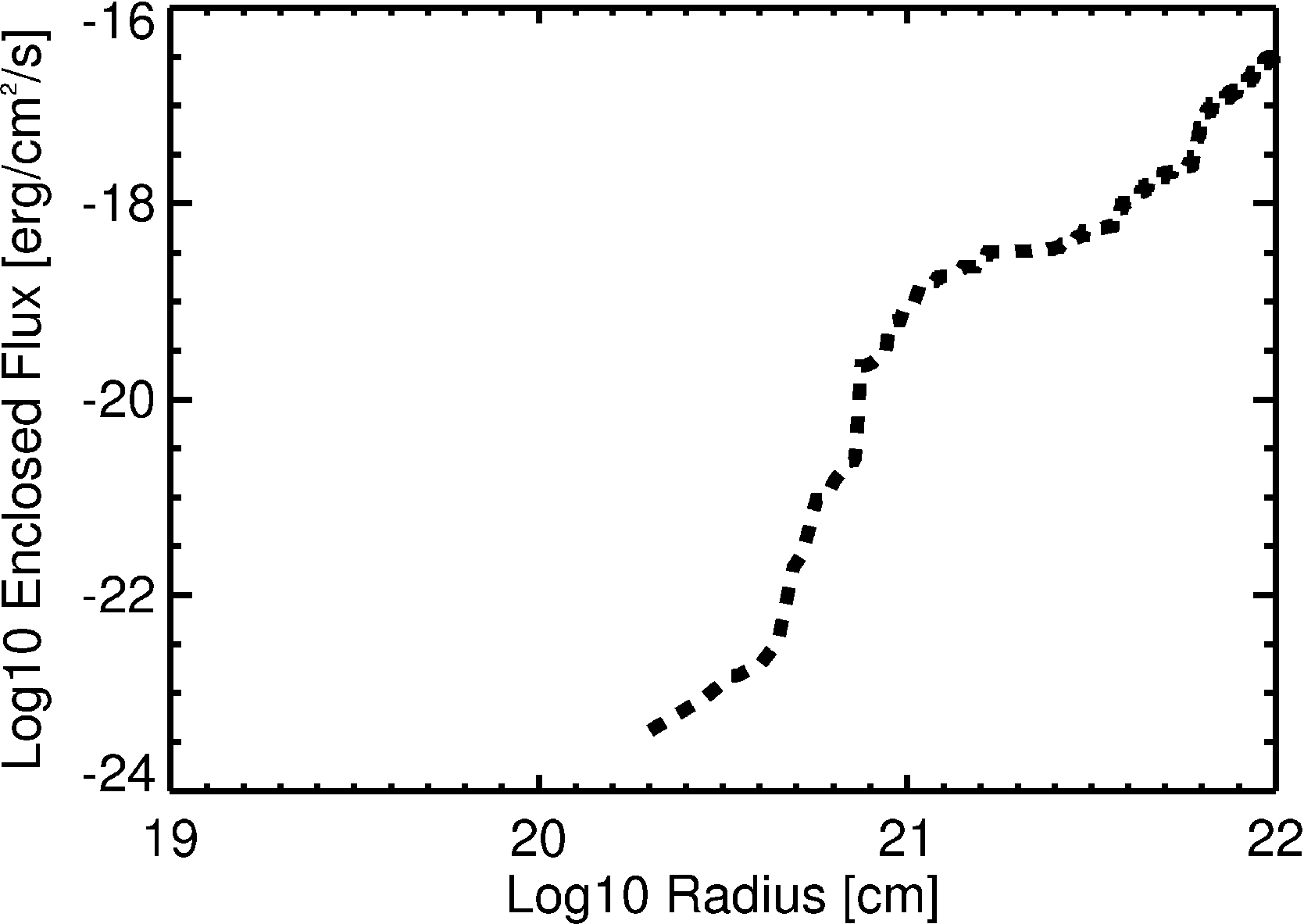}
\end{minipage}

\end{tabular}
\caption{All panels in this figure are log log plots. The upper left panel shows column density plotted against the radius of the halo. Lyman alpha emissivity radial profile emerging from the halo is shown in the upper right panel. Lyman alpha luminosity is shown in the lower left panel. Lyman alpha flux is plotted against the radius of the halo in the lower right panel.}
\label{figure3}
\end{figure*}

The initial density perturbations decouple from the Hubble flow and begin to collapse via gravitational instabilities. The gas is shock-heated during the non-linear collapse of density perturbations. At redshift 10, the halo begins to virialize. In the process of virialization, part of its gravitational energy is converted into thermal energy. This heats up the gas and results in the emission of Lyman alpha photons. The gas is accreted into the center of the halo through cold streams of $\rm \sim 10^{4}$K as shown in figure \ref{figure1}. The upper panels of figure \ref{figure1} show the density at different redshifts while lower panels show corresponding temperatures. Typical number densities of the cold streams are of the order of 0.01-1 $\rm cm^{-3}$. Density and temperature radial profiles are depicted in the upper left and right panels of figure \ref{figure2}. We see that in the presence of hydrogen line emission the gas cools isothermally. The density profile of the halo is $\rm \sim r^{-2.3}$, which agrees with results of \cite{2008ApJ...682..745W}. The ionization degree of the gas is shown in the bottom left panel (HII abundance) of figure \ref{figure2}. We see that most of the gas remains neutral at $\rm 10^{4}$ K, ionized fraction goes down due to faster recombination at higher densities $\sim n^{-0.5}$. The gas radial velocity is depicted in the bottom right panel of figure \ref{figure2}. It can be seen from the figure that gas is falling into the center of the halo. The velocity profile also shows that gas is accreted onto the halo through accretion shocks. The cold streams have typical column densities of about $\rm 10^{19} cm^{-2}$. A radial profile of the column density is shown in figure \ref{figure3}. At columns above $\rm 10^{21} cm^{-2}$ the gas optical depth exceeds $10^{7}$ and Lyman alpha trapping becomes effective. The Lyman alpha emissivity is shown in figure \ref{figure3}. It can be seen from the figure that above columns of the order of $\rm 10^{22} cm^{-2}$, Lyman alpha photons are trapped and cooling through this line is completely suppressed. That is why the Lyman alpha emissivity sharply declines in figure \ref{figure3}. We explored the local density variations and found that the maximum variation of the column density does not exceed an order of magnitude. In this sense, our calculation should provide a conservative lower limit on the expected flux. In the presence of Lyman alpha trapping, cooling can proceed through higher electronic states of atomic hydrogen \citep{2010ApJ...712L..69S}. For comparison, the total emissivity from higher states of atomic hydrogen and recombination/Bremsstrahlung processes is shown in figure \ref{figure3}. The total emissivity plot shows that despite line trapping, cooling still proceeds through these higher states of atomic hydrogen, particularly through 2s-1s and 3-2 transitions, and the thermal evolution is approximately isothermal.

The radial profile of the enclosed Lyman alpha luminosity is shown in figure \ref{figure3}. If the emission originates preferentially from the center, an approximately flat profile would be expected, while for uniform emission, a power-law behavior as $\sim r^3$ would be expected due to volume effects. Here, we find the absence of emission in the central core, then a sharp increase in luminosity between radii of $\rm 10^{20.2}cm -10^{21} cm$ by four orders of magnitude, and a more modest increase between radii of $\rm 10^{21} cm -10^{22} cm$. This behavior reflects the generation of Lyman alpha emission through accretion shocks. JWST can confirm the presence of such a brightness profile. The total Lyman alpha luminosity from the halo is $\rm 10^{44} erg~s^{-1}$, which is consistent with observed Lyman alpha blobs \citep{2009ApJ...696.1164O,2010MNRAS.407..613G}.

The emerging flux from the halo is depicted in figure \ref{figure3}. We find a total flux of $\rm 5.0\times 10^{-17} erg~cm^{-2}~s^{-1}$. At redshift 4.7, the observed Lyman alpha wavelength is at 0.68 microns. This can be detected with the JWST-instrument NIRSpec for integration times of $10^{4}s$, with S/N=$10$ and R=$100$. JWST spectrograph NIRSpec will have angular resolution of 0.1 arc-sec for 2 micron, and NIRcam will be well-suited for higher angular resolution with field of view of 2.2x2.2 $\rm arcmin^{2}~and~ 4096^{2}$ pixels for shorter wavelength of 0.6-2.3 micro meter. The halo in our case will have an angular size of 0.5 arc-sec at redshift 4.7. It should be detectable with NIRcam/NIRSpec. For extended emission of Lyman alpha, it may need higher integration time ($\sim 10^{5}$ sec) to resolve the flux from extended sources. The total mass of our halo at redshift 4.7 is $\rm 2\times10^{10} M_{\odot}$, which is not extreme in any way. Selecting higher mass halos will produce higher fluxes as there is a power law relation between the mass of a halo and its luminosity $\rm L_{Ly\alpha}\propto M^{5/3}$ \citep{2009Natur.457..451D,2006ApJ...649...14D}. We stopped our simulation at a redshift of 4.7 as it becomes computationally too expensive to follow the further time evolution.

In this work, we assume that there is no X-ray/UV background flux. Its presence can heat the gas and may increase the Lyman alpha flux \citep{2008ApJ...678L...5S,2007MNRAS.375.1269Z}. We have compared our results with \citep{2006ApJ...649...14D} and found good agreement. We have assumed here that the halo is metal free. The addition of small amounts of dust can absorb Lyman alpha photons efficiently, unless the gas is inhomogeneous \citep{1999ApJ...518..138H}. However, in-falling gas may still be pristine for isolated halos. We have ignored $\rm H_{2}$ cooling, which is suppressed in the presence of a strong UV background ($\rm J_{21}> 100$) \citep{2004ApJ...601..666D,2009A&A...496..365C}. \cite{2010MNRAS.402.1249S} found that even $\rm J_{21} \sim 30$ will be sufficient to quench $\rm H_{2}$ formation for local variations in flux see \cite{2008MNRAS.391.1961D}. At higher redshifts, it is possible that $\rm H_{2}$ may form for more modest radiation backgrounds. Even then, we expect H$_2$ formation preferentially at the center of the halo, while the accretion at the virial radius may still include a gas phase at $10^4$~K. This supports our main conclusion that Lyman alpha radiation originates mostly from gas in halo envelope.

Some LABs are also  associated with massive star forming galaxies \citep{2006ApJ...640L.123M} where stellar feedback or an AGN could power Lyman alpha radiation. It is clear that starbursts and AGNs are also potential candidates to power or enhance the observed Lyman alpha luminosities. It is not fully clear if our results are also applicable to such situations, although we expect similar effects of line trapping in the centers of such galaxies if $\rm N_{H}/\Delta v \geq 10^{21} cm^{-2} km^{-1} s$,  with $\rm \Delta v$ the cumulative velocity difference along the atomic hydrogen column $\rm N_{H}$.Current modeling uncertainties concern both the star formation efficiency and the initial mass function. Both radiative feedback, leading to the formation of HII regions, as well as mechanical feedback, providing a more clumpy structure with higher escape fractions, need to be modeled self-consistently. Outflows if present may enhance the detectability of Lyman alpha emission in the outskirt of the halo,  5 \% of the emitted Lyman alpha photons could be directly transmitted to the observer along the line of sight \citep{2010MNRAS.408..352D}. In the future, cosmological radiative transfer simulations including feedback effects should be performed to obtain more robust results.


\section*{Acknowledgments}
The FLASH code was in part developed by the DOE-supported Alliance Center for Astrophysical Thermonuclear Flashes (ACS) at the University of Chicago. DRGS acknowledges funding via the European Community’s Seventh Framework Programme (FP7/2007-2013) under grant agreement No. 229517 and via HPC-EUROPA2 (project number: 228398) with the support of the European Commission Capacities Area–Research Infrastructures Initiative. SZ thanks the ladt Davis foundation for support. We thank the anonymous referee for a careful reading of the manuscript and many insightful comments.

\label{lastpage}

 \bibliography{biblio3.bib}

\end{document}